# Enhanced capacity & coverage by Wi-Fi LTE Integration

Jonathan Ling, Satish Kanugovi, Subramanian Vasudevan, A Krishna Pramod

### Abstract

Wi-Fi provides cost-effective data capacity at hotspots in conjunction with broadband cellular networks. The hotspots are required to capture a large number of users and provide high data rates. Data rates, over the Wi-Fi interface, are influenced by the media access protocol, which loses throughput due to delays and unintended collisions when a large number of users are active. The hotspot range which determines the number of users, that can associate, is limited by the lower power of the client rather than the access point. By diverting the traffic destined to the access point via another access network, both range and efficiency can be improved. This uplink redirection or diversion is achieved by simultaneous use of the Wi-Fi and LTE radio interfaces. Three options - loose, tight, and hybrid integration are presented towards providing enhanced capacity and coverage.

### Introduction

Large scale adoption of smart phones and tablets means that wireless data networks must provide high data rates anytime and anywhere. Wi-Fi networks are characterized by high peak rates, but lower efficiency for small packets, and by limited coverage. 3GPP LTE networks on the other hand are characterized by ubiquitous coverage and high spectral efficiency. LTE small cells are beginning to be deployed as a complement to LTE macro networks to address the need for higher cellular capacity. To utilize unlicensed spectrum allocated at 2.4 and 5 GHz, the LTE network may also be augmented with Wi-Fi access points (APs) integrated with the LTE small cells or deployed separately in their neighborhood. Optimizing the data traffic over the two interfaces benefits the user as well as the service provider. For example, serving best effort traffic over the Wi-Fi interface and QoS sensitive traffic over the LTE interface enhances user experience in both traffic classes by leveraging the large bandwidth available on the Wi-Fi link and freeing up resources on the LTE interface.

In Wi-Fi/LTE heterogeneous networks, current schemes for integration [1] allow users access to one or the other, or concurrent use of both radio interfaces, but for different applications [1][2]. For example, a web browsing user may be switched from LTE to Wi-Fi access when Wi-Fi becomes available. The heterogeneous network may support both low bandwidth, low latency applications such as voice-over-IP on LTE and web-browsing on Wi-Fi in a region of simultaneous LTE and Wi-Fi coverage.

Management of switching between the access networks is currently device dependent or network assisted using policies communicated to the user. Policies are stored in an access network discovery and selection function (ANDSF), in the core network and communicated to the client's device. The client device is responsible for execution of the policy, i.e. selection of the appropriate air interface, based on the policy received from the network. Standardization efforts are ongoing in 3GPP [3] to further supplement ANDSF based policies with radio access network (RAN) information.

The evolution of these mechanisms, as shown in Figure 1, will be towards the simultaneous use of both the interfaces by an application in order to further reduce service delays and consistently maintain user experience. Experience on the Wi-Fi network can be uneven due to excessive loading [4]-[6]. Due the imperfect nature of radio channel sensing, as well due to contention based access, packet collisions can occur. Collisions can occur between all co-channel transmitters: access points (APs) on the downlink, APs transmitting and the client transmitting, and between clients transmitting on the uplink. The user may also move in and out of hotspot coverage. Overall coverage is limited by the relatively low power of the Wi-Fi client. As the Wi-Fi and LTE networks have different loads, capabilities, and characteristics, joining and optimizing over the two is expected to improve user experience.

By evolving to simultaneous use of LTE and Wi-Fi interfaces, we attempt to provide gains beyond simple federation bandwidth. This has been called super-aggregation being that the benefits are greater than the sum of their components [7]. The capacity of a Wi-Fi hotspot may be improved by redirecting uplink traffic from the devices (that are capable) to the LTE radio interface and thus removing the source of contention and collisions.

In the sections that follow as both background and motivation we describe aspects of Wi-Fi capacity and coverage. We present three levels of network integration: (1) loose integration using a high layer control protocol that interacts with the application and both radio interfaces (2) tight integration with Wi-Fi and where the optimizing function is part of the LTE radio link controller (3) hybrid integration where the LTE network provides an alternative route back to the WiFi AP. Assuming one of the forms of super-aggregation is present; we provide some additional thoughts on system optimization.

Subramanian Vasudevan (subramanian.vasudevan@alcatel-lucent.com), Satish Kanugovi (satish.k@alcatel-lucent.com), Jonathan Ling (Jonathan.Ling@alcatel-lucent.com), A Krishna Pramod (krishna_pramod.a@alcatel-lucent.com) are at Alcatel Lucent.

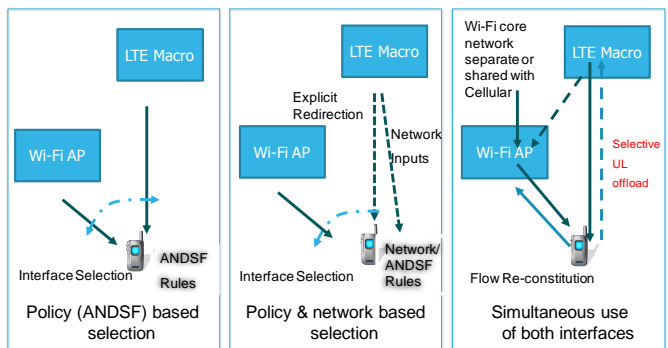

Figure 1 Evolution of Wi-Fi LTE integration schemes.

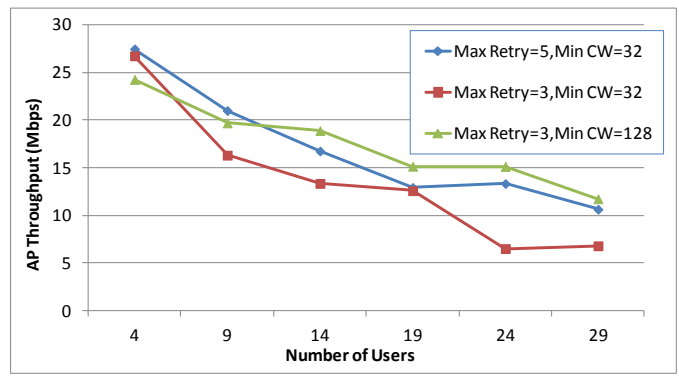

Figure 2 Degradation of Wi-Fi total throughput with increasing number of users for different operating parameters.

### WI-FI CAPACITY

Uplink and downlink transmissions time-share the wireless channel using a polite media access control (MAC) protocol called Carrier Sense Multiple Access with Collision Avoidance (CSMA/CA) [9]. According to the protocol, each potential transmitter, either an access point or client, must defer transmission until the shared channel is deemed to be clear. CSMA/CA not only prevents collisions, but allows Wi-Fi to be robust and scalable for large planned and unplanned/ad-hoc deployments.

CSMA/CA however has been reported to be the cause of performance degradation for large number of users and dense deployments [4]-[6][10]. We have observed similar trends in simulation for increasing number of clients. Using the event driven simulator NS-3, experiments were run where clients attempt to send as much traffic as possible on both uplink and downlink. That is, there is always data available to send. As the number of clients is increased the total throughput degrades due to packet collisions, as shown in Figure 2. The simulator modeled 802.11a clients at fixed radio transmission rate of 54 Mbps, and was modified from capturing packets on a first-come first serve basis, to a realistic model that captures packets based on signal-to-noise and interference ratio (SINR). In the experiment we also varied the contention window size and the maximum number of packet retries. Increasing the contention window increases the overhead but reduces the probability of collision. Increasing the number of retries reduces the probability that a packet will be lost, and thus need to be transmitted by a higher layer protocol. Increasing the contention window resulted in lower throughput for small number of clients and an improvement at high number of clients with high collisions. Adjustments to both parameters however, do not truly mitigate the drop in throughput especially when one considers single client throughput as the baseline.

There are also efforts within the Wi-Fi standards groups to enhance throughput and performance. The IEEE standard 802.11ac provides higher bandwidths to individual clients by bonding across available channels [11]. Since channels are taken away from other users, 802.11ac provides full benefits in sparse networks. Meanwhile the IEEE 802.11 HEW group is addressing dense networks. The Dual Wi-Fi [5] proposal requires carving out blocks of spectrum for separating uplink and downlink from the common Wi-Fi channel based on relative uplink/downlink load. While this solves the issue of contention between uplink and downlink and enhances spectral efficiency, it reduces the available bandwidth for either of the links since a portion of it has to be avoided for exclusive use by the other link.

### WIFI COVERAGE

Figure 3 illustrates how coverage differs between Wi-Fi and LTE and the impact it has by creating areas of different capacity. Consider a heterogeneous network deployment where macro cell coverage is provided by LTE with an embedded small cell providing service using both LTE and Wi-Fi. The LTE coverage represents a wide area across which capacity can be distributed amongst large number of users. The LTE small cell provides higher data rates to users than the LTE macrocell due to better channel quality at the receiver due to the proximity of the user to the small cell. The Wi-Fi cell provides additional capacity but has limited range over which users receive benefits.

In Wi-Fi networks, the range is usually determined by the client with lower uplink transmit power, rather than the AP. Clients typically transmit at 40 mW due to cost and safety reasons whereas APs often have the capability to transmit up to maximum regulatory power, e.g. 4W EIRP at 2.4 GHz in the United States. Given the higher downlink power, and if not constrained by the uplink, range can be increased by about $3\times$ and area coverage by $10\times$ according to typical distance dependence of inverse $4^{th}$ power. Note for cellular networks such large asymmetry is typically not present. This is because the larger transmit power at the base station is divided among all the users, which are served concurrently

Range has direct impact on a large network's deployment costs. Indoor APs are lower cost because the building or home owner provides the backhaul and electricity, and the APs themselves have no need for environmental hardening. Radio signals are strongly attenuated by exterior walls, and thus signals from indoor transmitters are received weakly outdoors [12]. Thus for full Wi-Fi coverage separate indoor and outdoor APs must to be installed [13].

By avoiding this uplink range limitation, the range of Wi-Fi and hence the area over which high data throughput can be provided can be increased. If Wi-Fi were used in one-way or downlink only mode then range could be increased, and a lower cost network with coverage outdoors using indoors APs or vice-versa would be possible. We describe the architecture towards achieving this in the upcoming section on integration.

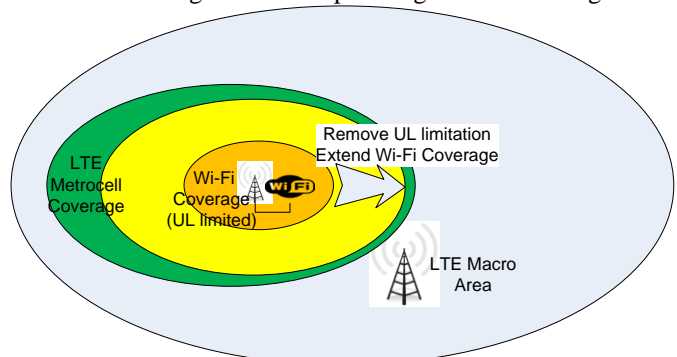

**Figure 3: Coverage differences in a Wi-Fi/LTE Heterogeneous Network.**

### INTEGRATION OF WI-FI AND LTE

In this article, we present solutions towards enhancing efficiency and range of Wi-Fi, by redirecting uplink traffic from the Wi-Fi network interface to the LTE network interface. Depending on the architectural variant some traffic may still be transported on the Wi-Fi uplink, e.g. management packets. Also in some variants downlink traffic may also be split between the Wi-Fi and LTE networks providing benefits of offload to the cellular network. The uplink traffic, being routed via the LTE interface, is completely scheduled and hence offers better performance to the user. When the AP only delivers downlink traffic there is no contention to resolve using the CSMA protocol and in a planned network with no co-channel neighbors, Wi-Fi can operate on completely scheduled basis. Downlink rates can be guaranteed since the channel use is fully under the control of the scheduler. Guaranteed rates can be extremely helpful in providing high quality video. Re-direction should occur when the contention due to uplink traffic on the Wi-Fi interface begins to impact channel availability. Where redirection of both management and data packets is achieved; extended coverage is possible since Wi-Fi will not be limited by the lower power of the client.

Three distinct Wi-Fi/LTE integrated solutions are presented. In loose integration network connections can be managed separately without knowledge of each other's presence. Loose integration requires only modification to operating software in the client device and server. It knows the least about the network state, provides some performance improvement, and is easily implementable. In tight integration both networks are closely coupled to potentially provide the highest performance, but at high implementation complexity. We also present a hybrid solution where networks are minimally aware of each other, and demonstrate how simple routing of packets over the alternate network can improve coverage. In terms of complexity, hybrid integration falls between loose and tight integration.

### LOOSE INTEGRATION

Multipath TCP (MPTCP) is an IETF draft standard that defines a method of joining independent TCP sub-streams to provide a reliable connection between client and server applications [8]. MPTCP provides bandwidth federation and redundancy. Redundancy can be used for to support user mobility over wired and wireless networks. The current specification provides control, scheduling, and compatibility with existing internet routers. Note in the TCP/IP networking reference model TCP and MPTCP operate on what is called the transport layer which communicates up to the application and downwards to the network layer that handles routing. Refer to [9] for additional details.

We consider augmenting MPTCP with a rule that specifies how an interface is used when it is available. The rule controls how traffic is optimized, and may also utilize statically configured information such as air-interface overhead for particular packet size. Figure 4 illustrates the architecture for loose integration using MPTCP. Two different modes of operation are shown - one on the left for achieving downlink traffic aggregation that can be used to offload downlink traffic from LTE to Wi-Fi and, the other mode on the right is used for offload of uplink traffic from Wi-Fi to LTE for improving efficiency on the Wi-Fi air interface.

By changing the uplink "cost" on the Wi-Fi TCP subflow, along with using existing metrics such as measured round-trip time and bandwidth, MPTCP would schedule the majority of the uplink traffic on the LTE TCP subflow. This would increase Wi-Fi efficiency, but not as much as if all uplink signaling were shifted. That is, on the Wi-Fi uplink TCP acknowledgements (ACKs) are still being generated and in response to downlink TCP packets. While shifting the ACKs to a different interface appears feasible, it does break the independence of TCP subflows. Finally, in theory, to improve responsiveness MPTCP could utilize cross layer information, e.g. Wi-Fi air-time usage, packet retry counters, and signal strength, but this comes at the cost of impacting the robustness and simplicity provided by layering network services and protocols.

### TIGHT INTEGRATION

In this architecture we consider tightly integrated and optimized networks. Tight network integration is specified in SaMOG [14], where the Wi-Fi AP is a trusted 3GPP/LTE network element. This is accomplished by 802.11i security and connecting Wi-Fi APs by secure tunnels to the Trusted Wireless LAN Gateway (TWAG). The connections between the elements are assumed to be high speed and reliable.

Figure 5 illustrates the architecture where traffic, also known as a bearer in 3GPP, is optimally distributed across the radio technologies via a control unit (CU) at the base station, called eNB in 3GPP. Traffic enters and exits the LTE RAN via the PGW. Reference [15] provides additional details on the 3GPP network architecture. As example use cases, for the tight integration solution, two modes of operation are shown in Figure 5 - the one of left demonstrates downlink bearer aggregation across Wi-Fi and LTE interfaces for providing cellular offload and the other mode on the right illustrates offload of uplink TCP traffic to LTE from Wi-Fi for enhanced Wi-Fi efficiency. Note in Figure 5, the path through the TWAG is not shown for clarity. The two key functional units of this system are a central control entity at the eNB that meets each of users' quality of experience (QoE) requirements by determining the split of the bearer between the Wi-Fi and LTE interfaces in the downlink and a control unit at the client device that handles aggregation and feedback for the downlink traffic and routes uplink traffic.

The existing radio link controller at the base station can be upgraded to handle both links. A similar control function operates at the client to distribute data across the radio interfaces. Both of these obtain feedback on the channel capacity and buffer drain rates from the existing network stack. Using this information, the control unit can quickly respond to dynamic RF conditions and user loads to tailor the bearer distribution between the Wi-Fi and LTE interfaces. Due to the direct availability of the feedback at the control unit, it can respond faster to the access link conditions as compared to the loose integration solution, where the transport layer has to infer the congestion in the links using secondary effects like round trip delays and TCP ACK loss. In this architecture, both the transport signaling, e.g. TCP ACKs and application data in the uplink can be transferred to the LTE air link. This is an advantage over loose (MPTCP based) integration where the TCP ACKs for the downlink data need to be sent via the interface.

While the loose and tight integration schemes provided enhanced Wi-Fi performance by offloading different components of the application and management traffic in the uplink, the Wi-Fi protocol based uplink control signaling such as MAC layer ACKs, probe and association requests for Wi-Fi link setup, need to be carried over the Wi-Fi uplink. The range at which the device can be served by the Wi-Fi AP is therefore, still, limited by the Wi-Fi uplink. Further, elimination of Wi-Fi protocol's control signaling, like ACKs from the Wi-Fi uplink can reduce interference caused to the neighboring APs and hence improve performance in dense deployments. A method to deal with this is explained in the hybrid integration solution.

### Hybrid integration

Instead of switching and disabling the alternative interface the connection manager may instead allow both interfaces to be active and available. Each will be independently assigned an IP address. Applications can be connected to one or the other interfaces, or even both. Note that the independent operation of IP path through the Wi-Fi interface is called non-seamless wireless offload (NSWO) in the 3GPP parlance [15].

In the hybrid integration solution, as illustrated in Figure 6, Wi-Fi interface is assisted by the LTE network to transport the uplink Wi-Fi packets (Uplink bearer like Application data and TCP Control signaling, uplink control and management frames like Wi-Fi MAC layer ACKs) via a tunnel to the Wi-Fi AP. As an illustration, say an application that uses only the IP interface via Wi-Fi, like App1 in Figure 6, the downlink data flows only over the Wi-Fi radio and uplink traffic is tunneled via the LTE interface to the AP. From the perspective of the AP, all uplink traffic appears to flow to and originate from the Wi-Fi interface. The tunnel traverses the LTE radio access network(RAN) through the eNB and terminates at the WiFi AP either through the LTE packet core, which is the regular path for any LTE bearer or a via a direct path between the eNB and the AP. While the direct path requires enhancements to the eNB, it offers the advantage of lesser latency over the tunneled link. APs terminate the IP tunnel by injecting the packets into the networking stack at a point where they would normally be obtained from the radio itself. Otherwise, both the networks remain independent and unaware of each other. As an illustration of the independence of the two interfaces, we have shown in Figure 6, another application on the device, App2 that uses the IP path via the LTE interface, operating independent of the hybrid mode of operation for App1 that uses LTE assisted Wi-Fi link.

The integrated Wi-Fi/LTE system, with all uplink transmissions (data and management) transferred from Wi-Fi to LTE, enhances the range and capacity of Wi-Fi. This makes it well suited for operation in dense deployments by reducing uplink interference to neighboring APs. When APs are sparsely placed the capacity of a heterogeneous network is enhanced by being able to serve users in a larger range and higher downlink capacity from the APs. The hybrid integration mode can be used in conjunction with loose and tight integration mode to achieve further path optimization by enabling TCP ACKs and Application bearer to directly reach their protocol end-points using the LTE interface without the need to traverse the tunnel via the AP, wherein only the Wi-Fi protocol control and management information uses the uplink tunnel to reach the AP.

### System Level Considerations

Existing multi-radio, Wi-Fi/LTE, clients can be enhanced by software upgrade to support the Wi-Fi/LTE super aggregation schemes (integrated Wi-Fi mode) discussed in the earlier sections. These clients will need to co-exist with existing standard multi-radio and single radio Wi-Fi devices. The integrated Wi-Fi mode may be attached to a separate service set identifier (SSID) to distinguish itself from standard access. That is, the Wi-Fi AP will broadcast at least two SSIDs: one for standard Wi-Fi access and the other for integrated access. For hybrid integration its SSID would be transmitted at full power and for standard access the SSID would be transmitted at a power close to that of to the client. When in range of both SSIDs the connection manager would determine whether to enable integrated mode or stay in the standard connection mode.

Overall Wi-Fi efficiency can be improved by designating short time intervals as downlink only periods. During this period downlink transmissions for integrated Wi-Fi and standard clients are scheduled. Since the newer devices in the integrated Wi-Fi mode have their uplink redirected to LTE, they can fully leverage the improved Wi-Fi performance without any negative impact on their uplink performance. This is an improvement over the standard Wi-Fi mode devices whose uplink will have to wait until the downlink-only period is completed. In order to reserve this period the network allocation vector (NAV) of nearby clients must be updated. The AP should broadcast a Clear To Send to Self (CTS-To-Self) 802.11 management message for the airtime reservation of up to 32 ms for the data in its buffer. In the integrated mode, contention is reduced therefore the contention window can be tuned to operate at the minimum levels to control the protocol overhead.

LTE systems can either be Frequency Division Duplex (FDD) or Time Division Duplex (TDD). In FDD LTE systems, separate frequency bands are reserved for uplink and downlink. In TDD LTE systems, the resources (time slots) on a common frequency band are shared between uplink and downlink, with the uplink to downlink resource ratio being tune-able based on relative traffic load. TDD LTE systems, therefore, offer further flexibility by allowing appropriate tuning of the uplink-downlink resource ratio to accommodate the additional uplink load offload from Wi-Fi to LTE.

## CONCLUSIONS

With the tremendous growth projected in the demand for data from the applications and devices leading to the hunger for more and more bandwidth, Wi-Fi provides a cost-effective way to address the high data capacity needs by leveraging the unlicensed spectrum. The limitation on Wi-Fi spectral efficiency and coverage due to the uplink, has been outlined, and it has motivated solutions that integrate Wi-Fi with LTE access networks. The expanding availability of LTE coverage with the ubiquitous presence of multi-radio devices offers different options for integration, with different levels of implementation complexity and the corresponding trade-off in benefits. These enhancements enable Wi-Fi networks to serve more users with higher throughput demands and as an effective traffic offload solution for cellular operators, all coming with the cost benefits of using the unlicensed spectrum.

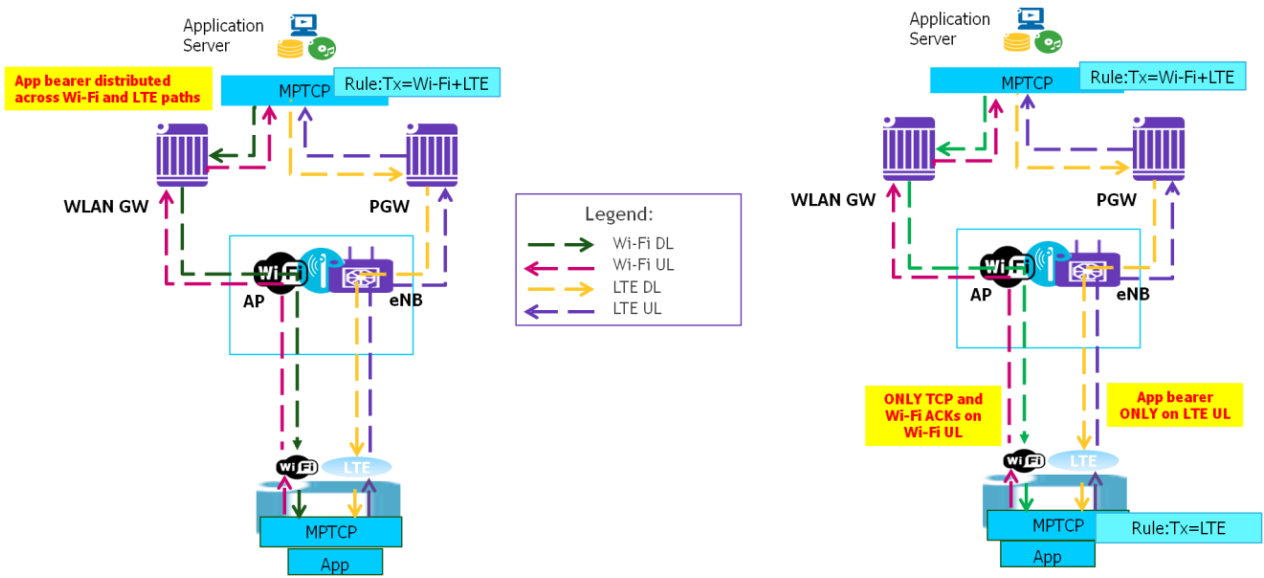

**Figure 4 Architecture for loose Wi-Fi LTE integration**

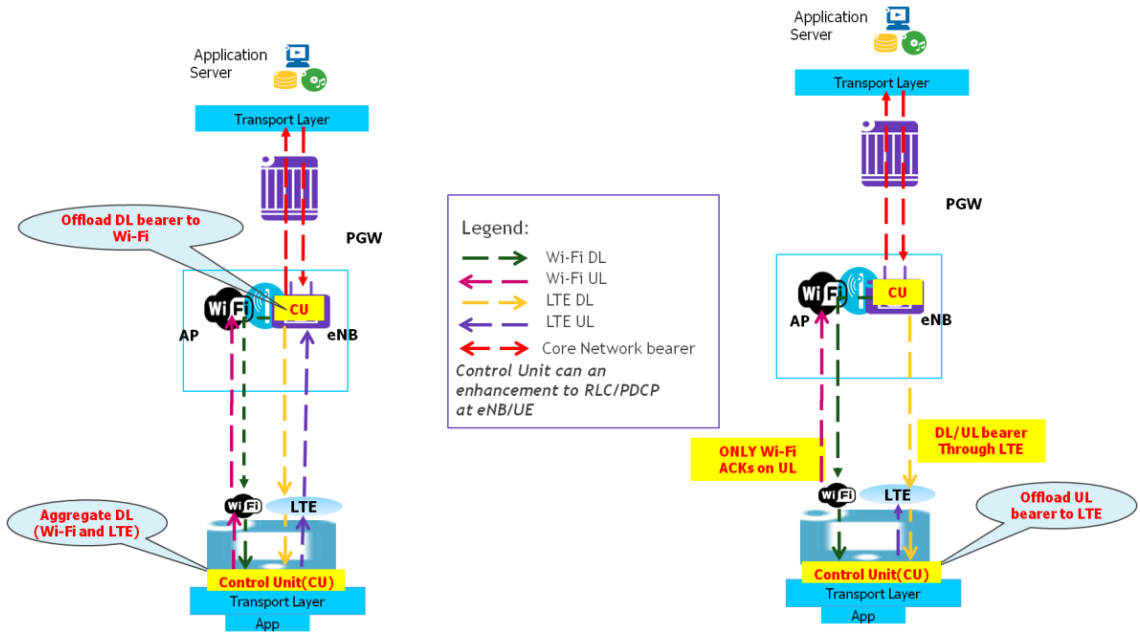

**Figure 5 Architecture for tight Wi-Fi LTE integration**

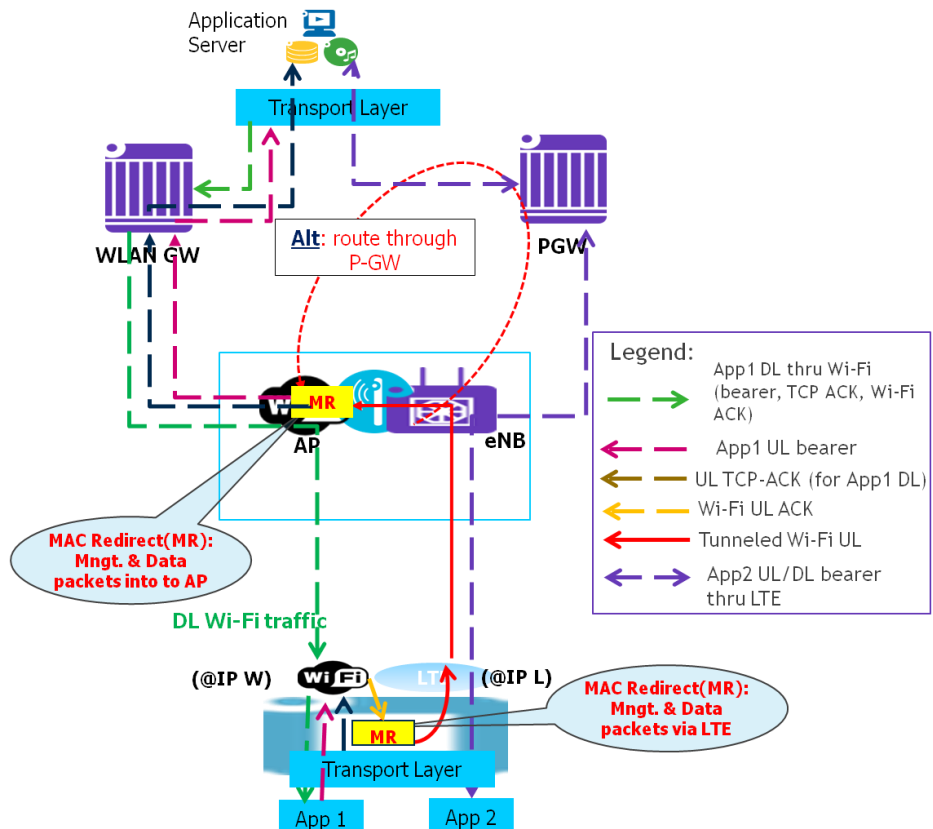

**Figure 6 : Architecture for hybrid integration. App1 is in hybrid mode. App2 uses LTE only.**